**Community-driven dispersal in an individual-based predator-prey model**

Elise Filotas*, Martin Grant†, Lael Parrott* § and Per Arne Rikvold‡

*Complex Systems Laboratory, Département de Géographie, Université de Montréal, C.P. 6128, Succursale Centre-ville, Montréal, Québec, H3C 3J7, Canada,

†Department of Physics, McGill University, 3600 rue University, Montréal, Québec, H3A 2T8, Canada

‡ School of Computational Science, Center for Materials Research and Technology, National High Magnetic Field Laboratory, and Department of Physics, Florida State University, Tallahassee, FL 32306, USA

§ Corresponding author

**email**: lael.parrott@umontreal.ca; **tel**: 1 (514) 343 8032; **fax**: 1 (514) 343 8008

**Abstract**

We present a spatial, individual-based predator-prey model in which dispersal is dependent on the local community. We determine species suitability to the biotic conditions of their local environment through a time and space varying fitness measure. Dispersal of individuals to nearby communities occurs whenever their fitness falls below a predefined tolerance threshold. The spatiotemporal dynamics of the model is described in terms of this threshold. We compare this dynamics with the one obtained through density-independent dispersal and find marked differences. In the community-driven scenario, the spatial correlations in the population density do not vary in a linear fashion as we increase the tolerance threshold. Instead we find the system to cross different dynamical regimes as the threshold is raised. Spatial patterns evolve from disordered, to scale-free complex patterns, to finally becoming well-organized domains. This model therefore predicts that natural populations, the dispersal strategies of which are likely to be influenced by their local environment, might be subject to complex spatiotemporal dynamics.

## 1. Introduction

During the past 20 years, spatial modeling has gained increased recognition as a theoretical tool to understand and study spatially structured populations (Hogeweg,1988; Bascompte and Solé, 1995; Hanski, 1998; Bascompte and Solé, 1998). Interest in such models has emerged in parallel with the desire to comprehend how space contributes to population dynamics (Hassell et al., 1994; Bascompte et al., 1997; Ranta et al., 1997; Bjørnstad et al., 1999; Blasius et al., 1999) and to achieve insights into the origin of the many spatiotemporal patterns observed in nature (Bascompte and Solé, 1998; Marquet, 2000; Wootton, 2001).

One central mechanism in spatially explicit models is species dispersal. Unfortunately, it has been difficult to establish regular and common rules governing species dispersal from the numerous empirical studies of individual movements between habitats. This absence of general behavioral rules has often brought theoretical ecologists to adopt the simplest possible assumptions when modeling dispersal processes (Bowler and Benton, 2005). Most spatial models have been designed using a density-independent rate of dispersal, which implies that a constant ratio of the local populations moves in each generation, regardless of local conditions (Solé et al., 1992; Hastings, 1993; Bascompte and Solé, 1994; Hassell et al., 1995; Rohani et al., 1996; Kean and Barlow, 2000; Kendall et al., 2000; Sherratt, 2001). This random or passive dispersal indeed operates on many groups of organisms (some invertebrates, fish, insects and sessile organisms such as plants) that depend on either animal vectors, wind or current for dispersal (Maguire, 1963; Bilton et al., 2001; Nathan, 2006). On the other hand, it is now well recognized that dispersal for many animals largely depends on factors such as local population size, resource competition, habitat quality, habitat size, etc (Johst and Brandl, 1997; Bowler and Benton, 2005). Therefore, recent models have started to incorporate more varied dispersal rules, and results suggest that the dispersal mechanism can significantly influence modeling predictions. One such dispersal rule, which has received great attention, is the use of a density-dependent rate of dispersal (Amarasekare, 1998; Ruxton, 1996; Doebeli and Ruxton, 1998; Sæther et al., 1999; Ylikarjula et al., 2000; Amarasekare, 2004). A positive rate expresses intra-specific competition, while a negative rate mimics the inconveniences associated with isolation, such as higher predation risk and foraging and mating difficulties. Other condition-dependent dispersal strategies have also been investigated, such as dependence on habitat saturation (South, 1999), the dependence on resource availability (Johst and Schöps, 2003), or migration following the theory of the ideal free distribution (Ranta and Kaitala, 2000; Jackson et al., 2004), to name but a few. For a thorough review see Bowler and Benton (2005). These studies focus on the effect of condition-dependent dispersal on the persistence of populations in space, as well as the stabilization and synchronization of their dynamics.

Here, we explore the effects of a novel community-driven dispersal strategy on the dynamics of spatially structured predator-prey populations using an individual-based model. We measure the impact of the community on its constituent species using a single quantity designed to take into account the effects of interspecific competition, intraspecific competition and resource availability on the individuals of the system. Hence, this quantity, which we name "fitness", is introduced as a way to quantify the multiple environmental pressures arising from various biotic factors that transcend simple population density. At this point it is important to clarify that the term fitness as used here does not have any evolutionary biology meaning (Ariew and Lewontin, 2004). The fitness of an individual, as used throughout this report, should not be confused with the usual definition of "expected number of offspring". We associate the term fit with the loose definition of a species being suited to a particular biotic environment and hence being apt to reproduce therein.

The dispersal strategy we adopt in our model encapsulates the idea that dispersal is a means for individuals to enhance their fitness. Here, the fitness of a species is a local quantity evolving in time, which influences the reproduction rate as well as dispersal. Individuals who are unfit to their community, relative to a predefined fitness tolerance threshold, are free to migrate in the "hope" of finding a more favorable habitat.

We study the spatiotemporal dynamics of this predator-prey model with respect to specific levels of tolerance through a quantitative analysis of the spatial patterns of correlation. We show that three distinct dynamical regimes emerge from this community-driven dispersal model, namely random motion, complex spatiotemporal patterns, and highly organized spatial domains. We also reveal that dynamics of such complexity cannot be generated with the use of density-independent motion.

## 2. Definition of the Model

We use an individual-based model inspired by the Tangled-Nature model (Christensen et al., 2002; Hall et al., 2002; di Collobiano and al., 2003; Jensen, 2004) and a similar model by Rikvold and co-authors (Rikvold and Zia, 2003; Zia and Rikvold, 2004; Sevim and Rikvold, 2005; Rikvold, 2006, 2007; Rikvold and Sevim, 2007), which are both non-spatial models of biological coevolution.  In these models, the individuals of the community are identified by their species genotype and interact via a set of fixed species-species interactions. Individuals reproduce asexually according to their fitness and are subject to mutation, which results in the creation of offspring of a different genotype. The fitness of a species varies with the type and strength of its interactions with the species of the community, as well as their respective population sizes. As the diversity and population sizes of species in the ecosystem fluctuate under reproduction and mutation, so does the fitness of each species.

Interest in such models comes from their simplicity and impressive intermittent dynamics over long time scales, which is reminiscent of punctuated equilibria (Eldredge and Gould, 1972). Lawson and Jensen (2006) have investigated the behavior of the Tangled-Nature model when coupled to a spatial lattice under density-dependent dispersal and found power-law species-area relations over evolutionary time scales. On the other hand, the focus of the present paper is the dispersal dynamics of a predator-prey system. We will therefore explore the behavior of this type of model on ecological time scales and with the addition of spatial degrees of freedom. To this end, we set the mutation rate to zero, and we associate the definition of fitness used in these coevolutionary models to the species reproduction probability, the local measure of a species' suitability to the local ecological community. Moreover, only two species are considered, a predator and its prey. A tacit benefit of using this framework obviously is that it could in the future be generalized to describe a multi-species system with mutation.

Space is modeled as a matrix composed of $LxL$ cells, each containing a non-spatial version of the model. Two processes control the time development of the model: reproduction, an intra-cell process, and dispersal, an inter-cell one. The probability that an individual of a given species $i$ reproduces is given by $f_i$, its species' fitness, where $i$ equals $v$ for the prey and $p$ for the predator. Reproducing individuals are replaced by two offspring, while individuals which are not able to reproduce are removed from the ecosystem. This procedure gives rise to non-overlapping generations. Dispersal, on the other hand, is controlled by the parameter $p_{motion}$, which has identical values for the predator and prey. Dispersing individuals travel to neighboring cells. After one reproduction and one dispersal attempt the process is reiterated. Note that migration is the only means of interaction between cells. Predators and prey are not allowed to feed from neighboring cells. The local population $n_i(x,y,t)$ of species $i$ at cell $(x,y)$ is therefore modified at each time iteration $t$, first through community-driven reproduction and second through community-driven dispersal.

### 2.1 The fitness

The fitness, $f_i$, quantifies how well species $i$ is adapted to its current community. The term fitness, as mentioned before, does not have any Darwinian meaning in this report. Fitness is a characteristic of an entire species and not of a single individual. $f_i$ also represents the reproduction probability of species $i$ and is defined over the interval [0,1]. A low fitness value implies that species $i$ lives under harsh biotic conditions and hence its probability of reproduction is low in this specific habitat. Conversely, when $f_i$ is large, species $i$ is suited to the local community and its reproduction probability is consequently high. The fitness of a species $i$ is given by (Rikvold, 2006, 2007; Rikvold and Sevim, 2007):

$$f_i(x,y,t) = \frac{1}{1+\exp[\Phi_i(x,y,t)]} \in [0,1] \tag{1}$$

where

$$\Phi_i(x,y,t) = \frac{1}{c}\left( \beta_i - \frac{R\eta_i/N(x,y,t)}{1+R\eta_i/N(x,y,t)} - \frac{1}{N(x,y,t)}\sum_j J_{ij} n_j(x,y,t) \right). \tag{2}$$

The function $\Phi_i(x,y,t)$ can be thought as measuring the impact of the local *(x,y)* community on species $i$ at time *t*. The parameters are defined as follows:

- $n_i(x,y,t)$ is the population size of species $i$ in the cell of coordinates *(x,y)* and at time *t*

- $N(x,y,t)$ is the population size of the community located in cell *(x,y)* at time *t*, i.e.
  $$N(x,y,t) = \sum_{i=v}^{p} n_i(x,y,t)$$

- $\beta_i$ is the cost of reproduction of species $i$; it is a real number between 0 and 1. The higher $\beta_i$ is, the harder it is for the species to reproduce.

- $\eta_i$ is the coupling of species $i$ to the external resource. It is also defined on the interval [0,1]. In our model, $\eta_i$ is non-zero only for the prey as the predator does not feed on the external resource.

- $J_{ij}$ are the species-species interactions. Their values range over the interval [-1,1]. The off-diagonal elements of the matrix $J_{ij}$ are anti-symmetric. $J_{ij} < 0$ means that $j$ has a negative effect on species $i$, and $J_{ij} > 0$ means that $j$ has a positive effect on species $i$. Elements on the diagonal $J_{ii}$ determine intraspecific interactions. Although we confine our study to predator-prey systems, this formulation also allows for various types and strengths of interaction such as mutualism and competition (Rikvold and Zia, 2003; Sevim and Rikvold, 2005).

- $R$ is the size of the external resource. We fix $R$ to the same value in every cell to represent a homogeneous landscape.

- $c$ is a scaling parameter inversely proportional to the species' sensitivity to local conditions. A large $c$ is associated with low fitness sensitivity. In that case, every individual has more or less the same fitness regardless of their species and community population sizes and of the values assigned to $\beta$, $\eta$ and $J$. See figure 1. On the other hand, a small $c$ will enhance the influence of these parameters and will create higher fitness variability locally between the species and also between populations at different lattice sites.

The fitness is therefore a time, space and species dependent quantity. Consequently, a species can have a low fitness (and hence a low reproduction rate) in one region of the lattice and a higher one some distance away, depending on the present spatial distribution of the populations. Note that the functional response for the prey is a ratio-dependent modification of the common Holling type $II$ (Abrams and Ginzburg, 2000; Getz, 1984). This form was chosen because of its simplicity and generality, but it will be shown later that the exact shape of the functional response does not affect the general behavior of the model.

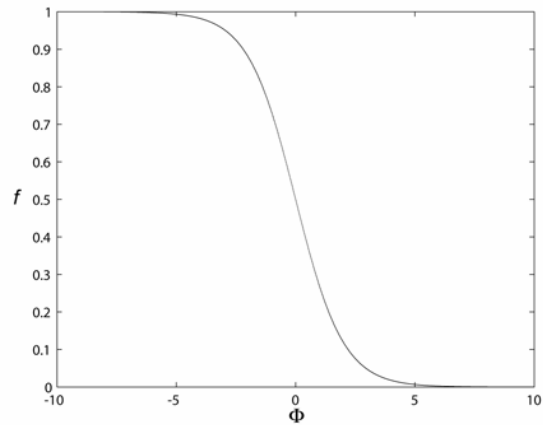

**Figure 1.** The fitness of a species as a function of $\Phi$.

### 2.2 The dispersal process

The dispersal rule in this model is motivated by the fact that dispersal is often seen as a means to improve an individual's condition. We hence allow individuals with low fitness to escape their site in the "hope" of finding a more suitable habitat. Following in philosophy the fitness-dispersal model of Ruxton and Rohani (1999), we set up a tolerance threshold called $p_{motion}$. An individual whose fitness is less than or equal to this threshold, $f_i(x,y,t) \le p_{motion}$, moves randomly to one of its neighboring cells. We choose a square neighborhood containing the individual's initial cell and the 8 immediately adjacent cells (also called a Moore neighborhood (Packard and Wolfram, 1985; Hogeweg, 1988)). Therefore, there is a 1/9 probability that an individual stays in its original habitat even when its fitness is less than the threshold $p_{motion}$. While it is possible that the displacement brings the individual to a more favorable environment, there is no guarantee that this goal is achieved. Even if the individual still has a fitness lower than $p_{motion}$ in its new community, it cannot disperse again. Contrary to purely density-dependent dispersal, this rule is clearly dependent upon the community as it does not simply depend on the migrating species' population density, but also on the local population size of the other species ($n_j(x,y,t)$) and on the resource availability ($R$). Note that while $p_{motion}$ has the same value for the predator and the prey, this does not imply that both species share the same dispersal rate, since the impact of the community on each species is different.

In order to appreciate how community-driven motion affects the predator-prey spatiotemporal dynamics, we compare the model to its density-independent dispersal version. In this classical scenario, the dispersal rule is controlled by the probability parameter $p_{ind} \in [0,1]$, which is a constant value independent of species, time and space. While it would seem ecologically more realistic to allow the predator and prey species to disperse at different rates, in this study, as it has been done elsewhere (Solé et al., 1992; Lawson and Jensen, 2006), we use a single parameter $p_{ind}$ for both species. This simplistic set-up permits a more direct comparison with the community-driven dispersal scenario. Therefore, this means that during each iteration of the model, each individual of both species located anywhere on the landscape has the same probability $p_{ind}$ to undergo dispersal, regardless of its fitness in the local community.

Therefore, for both models, the temporal dynamics of the local population $n_i(x,y,t)$ of species $i$ at cell $(x,y)$ is determined first by the community-driven reproduction process, where individuals are removed from the community and replaced or not by two offspring according to the fitness of their species, and second by the dispersal process, which controls the migration flow in and out of cell *(x,y)*.

## 3. Methods

### 3.1 Simulation details

The spatiotemporal dynamics of the model is investigated as a function of the dispersal parameters $p_{motion}$ and $p_{ind}$, for the community-driven and density-independent models respectively. The analysis is pursued by varying $p_{motion}$ and $p_{ind}$ from their lowest to their maximum value (0 to 1). Because the focus of this article is the consequence of community-driven dispersal, we fix the other parameters of the model:

$$R = 200 \quad \beta = \begin{pmatrix} 0.3 \\ 0.5 \end{pmatrix} \quad \eta = \begin{pmatrix} 0.5 \\ 0.0 \end{pmatrix} \quad J = \begin{pmatrix} -0.1 & -0.9 \\ 0.9 & -0.1 \end{pmatrix} \tag{3}$$

where the first coordinate corresponds to prey attributes *v* while the second coordinate is for the predator *p*. The parameters are chosen so as to generate an oscillatory predator-prey dynamics, but other selections could have been considered. Notice that we have selected a smaller cost of reproduction for the prey ($\beta_v = 0.3$) as generally prey have smaller body size than their predator and hence require less energy to reproduce. Moreover, we have set the predator-prey interaction to a high value ($\left| J_{vp} \right| = \left| J_{pv} \right| = 0.9$) to clearly express the negative impact of the predator on the prey and the converse positive impact of the prey on the predator.

We are also interested in the effect of the scaling parameter $c$ on the spatiotemporal dynamics of the system. The majority of our study is performed with $c$ fixed to 0.06, a choice based on the following ecological considerations. First, we require the fitness of a single prey individual in the absence of predators to be near unity and, second, the fitness of a single predator individual in the absence of preys should be near zero. While any $c$ smaller than 0.06 satisfies these two conditions, such values inconveniently produce a fitness which changes abruptly under small modifications of the population sizes. Indeed, due to the shape of the fitness curve as a function of $\Phi$ (figure 1), smaller values of $c$ generate fitness values that are mainly distributed on the top (fitness close to one) and bottom (fitness close to zero) branches of the curve with few intermediate fitness values between zero and one. On the other hand, with the intermediate value $c = 0.06$, the fitness of both species has a realistic sensitivity and can cover the entire range [0,1], offering enough variation to generate a rich and diverse dynamics. A detailed analysis of this scenario is pursued with simulations repeated over 100 different initial conditions. An initial condition corresponds to a random spatial distribution of the population, where predator and prey local populations $n_i(x,y,t)$ can take any value between 0 and 200 individuals. In addition, in order to explore how the dynamics fluctuates under fitness sensitivity to the community, we run a smaller number of simulations (20) with other values of $c$ chosen from the interval [0.01,0.4].

The simulations are carried out on a square lattice of side $L=128$ with periodic boundary conditions. Every run lasts 2048 generations. Although the system reaches a statistically stable state generally around 100 iterations, the results presented throughout this article are computed on the last 1024 time steps of the simulations.

During the simulations, we record the temporal evolution of four variables: the average species density and the local species density for each of the two species. The average density $\rho_i(t)$ is a global measure (equation 4). It is computed by counting the population size of species $i$ over the entire territory and normalizing by the total number of cells, $L^2$. The local density $D_i(x,y,t)$, on the other hand, is computed for each cell as the ratio of the local population size of species $i$ compared to the population size of the local community in that cell (equation 5).

$$\rho_i(t) = \frac{1}{L^2} \sum_{x=1}^{L} \sum_{y=1}^{L} n_i(x,y,t) \tag{4}$$

$$D_i(x,y,t) = \begin{cases} \dfrac{n_i(x,y,t)}{N(x,y,t)} & \text{if } N(x,y,t) \neq 0 \\ \\ 0 & \text{if } N(x,y,t) = 0 \end{cases} \tag{5}$$

### 3.2 Spatial pattern analysis: the structure factor

Previous studies have mainly adopted tools from non-linear dynamics, such as bifurcation graphs and Lyapunov exponents, when analyzing the outcome of their spatial models (Bascompte and Solé, 1994; Doebeli and Ruxton, 1998). Although these methods are useful to identify the presence of chaotic or complex regimes, they do not provide information concerning the spatial structures and the scales of emerging patterns. While the patterns of spatial synchrony produced by population models have been investigated, these analyses consisted in computing the temporal correlation between time series of the population density at different locations in the landscape (Ranta et al., 1997; Kendall et al., 2000). Such analyses do not provide information about the characteristic spatial scales of patterns on the landscape. Moreover, many studies of models that generate interesting complex structures provide only qualitative descriptions (Hassell et al., 1995; Li et al., 2005). This prevents any detailed comparison to be drawn between models varying in their dispersal strategy.

Here we employ two-dimensional spectral analysis to characterize the model's spatial structure. More precisely, we make use of the structure factor, akin to the R-spectrum, a standard method of analysis in condensed matter physics (Goldenfeld, 1992; Reichl, 1998) and statistical spatial ecology (Platt and Denman, 1975; Renshaw and Ford, 1984). The structure factor is simply the Fourier transform of the spatial autocorrelation function. It gives information about spatial patterns in reciprocal space instead of real space. The structure factor can be understood as the spatial analogue of the power spectrum of a temporal series of data. The power spectrum retrieves the frequencies at which a process varies in time. Similarly, the structure factor finds the wave numbers characterizing the spatial patterns. Just as the period at which a process repeats itself can be obtained as the inverse of the frequency for a time series, the inverse of the wave number gives the length scale of the spatial patterns. Therefore, the structure factor is a convenient tool when spatial structures have synchronous behavior. Moreover, the structure factor is computationally more efficient than the spatial autocorrelation function. Indeed, it is easily computed using the Fast Fourier Transform (FFT). For a landscape of size $LxL$, this algorithm necessitates only $L^2 \log L^2$ operations compared to the $L^4$ used to obtain the spatial autocorrelation function (Press et al., 1992). This difference is important when dealing with large territories.

The structure factor is defined in the following way. Let $D_i(x,y,t)$ be the density of species $i$ at cell $(x,y)$ and at time $t$. The two-dimensional Fourier transform of this quantity is:

$$\hat{D}_i(k_x,k_y,t) = \sum_{x=1}^{L}\sum_{y=1}^{L} e^{j2\pi k_x \frac{x}{L}} e^{j2\pi k_y \frac{y}{L}} D_i(x,y,t) \qquad (6)$$

where $j$ is the imaginary unit. The structure factor is the squared amplitude of $\hat{D}_i(k_x,k_y,t)$, averaged over a large number of initial conditions:

$$S_i(k_x,k_y,t) = \left\langle \left| \hat{D}_i(k_x,k_y,t) \right|^2 \right\rangle \qquad (7)$$

Because $D_i(x,y,t)$ is isotropic (i.e., patterns in our model do not favor any specific orientation), the structure factor can be averaged over the radial wave number $k = \sqrt{k_x^2 + k_y^2}$. Moreover, once the initial period is over, the structure factor remains statistically similar in time. We can thus average $S$ over time and write:

$$S_i(k) = \left\langle \left| \hat{D}_i(k) \right|^2 \right\rangle \qquad (8)$$

Our investigation of the spatial correlations in population densities will hence consist in the analysis and comparison of the structure factors calculated for each simulation under the variation of $p_{motion}$, for community-driven dispersal, and $p_{ind}$, for density-independent dispersal.

## 4. Spatiotemporal dynamics

### 4.1 Community-driven dispersal

In this section, we describe the spatiotemporal dynamics of the model when the dispersal depends on the local community. As will become clear in the next section, this dynamics is markedly different from the one emerging from traditional density-independent dispersal. The dynamics passes through three different regimes as $p_{motion}$ varies from 0 (no dispersal) to 1 (total dispersal), giving the appearance of phase transitions, although at this point we cannot say if they are true phase transitions or just smooth crossovers. Each of these regimes is characterized by specific spatiotemporal patterns, going from disordered to complex to highly organized domains, which can be categorized distinctively by their respective structure factors.

### Spatial analysis

We provide here a detailed description of the three dynamical regimes for the scenario $c$=0.06 (figure 2a). Regime $I$, of disordered patterns, corresponds to low fitness threshold values. Only individuals with very low fitness are allowed to disperse, and, as a consequence, few movements happen in the landscape. This dynamics gives rise to random patterns in the spatial population density: dispersal is not high enough to induce correlations between the populations of neighboring cells (figure 2b-c). The calculation of the structure factor confirms this absence of spatial structure. In figure 2e the structure factor plotted on a log-log scale depends only weakly on the wavenumber $k$.

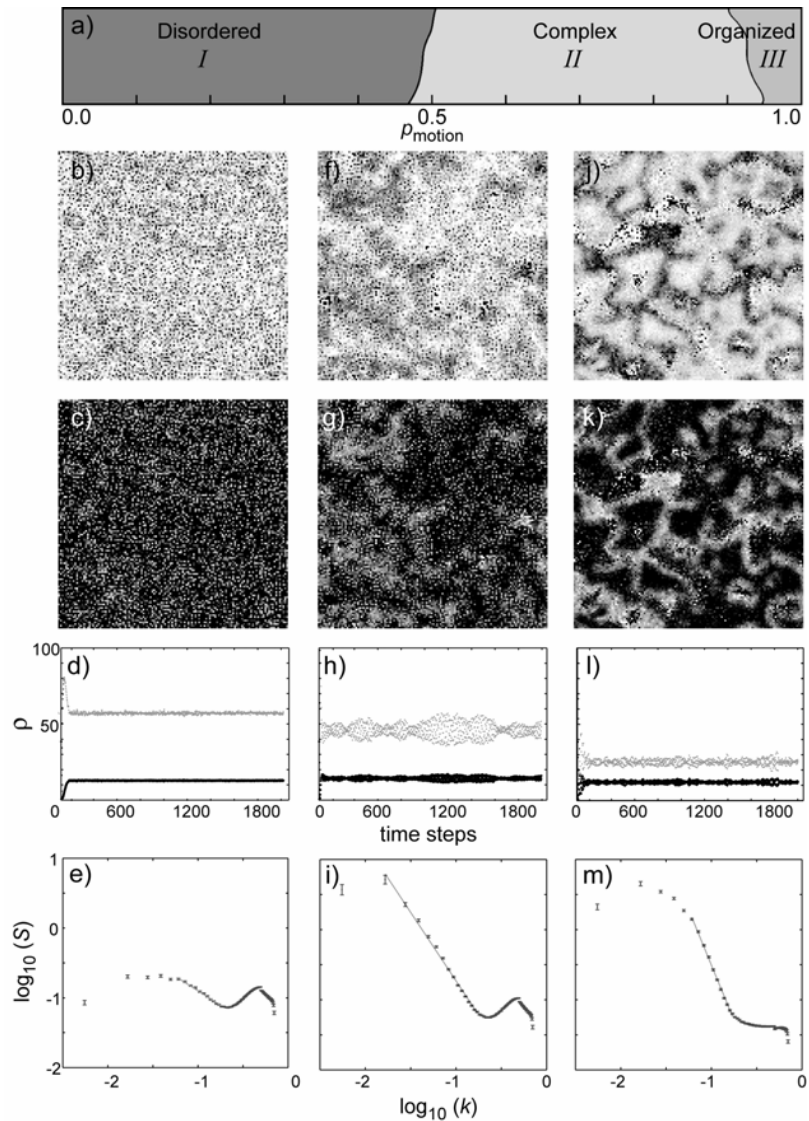

**Figure 2.** a.) Schematic representation of the three dynamical phases of community-driven dispersal: *I* - disordered, *II* - complex and *III* - organized domains (equivalent to figure 5a between $c = 0.05$ and $c = 0.1$). b.) Snapshot of the spatial prey density $D_v$ at $p_{motion} = 0.3$. c.) Snapshot of the spatial predator density $D_p$ at $p_{motion} = 0.3$. d.) Temporal evolution of the average density for the prey $\rho_v$ (gray) and for the predator $\rho_p$ (black) at $p_{motion} = 0.3$. e.) Log-log plot of the structure factor of the spatial prey density at $p_{motion} = 0.3$ (the structure factor for the predator is not shown since it is very similar to that of the prey). (f-g-h-i) same as (b-c-d-e) but at $p_{motion} = 0.7$. (j-k-l-m) same as (b-c-d-e) but at $p_{motion} = 1.0$. All figures are obtained with $c = 0.06$.

One can note however a peak of higher correlation around $k^*{\sim}0.5$. This indicates that population densities have similar values on cells at a small separation, $1/k^*$. This peak is caused by the short-range interactions between the neighboring cells. Consistently, the peak emerges at the same wavenumber for every simulation in regime $I$ and $II$, regardless of the value of $p_{\text{motion}}$. We have also confirmed (not shown) that as we change our definition of the cell's neighborhood, for example by enlarging it from 9 to 25 cells, the position of the peak also varies in inverse proportion. On the other hand, the other features of the structure factor are unchanged by the modification of the neighborhood's definition. Because it is not related to the general properties of the community-driven dispersal system, we can consider this peak to be an artifact of the model's construction. This peak indeed arises due to the grid representation of the landscape, which does not exist in natural systems. Therefore, it is a weakness of the model in the sense that the model best describes the general properties of the system only on large length scales ($k << k^*$) and not on the scale of single grid cells.

The complex regime $II$ is associated with intermediate fitness threshold values. Inside this regime the global population is very well divided between local populations of fitness below and above the threshold value. The motion of individuals in the landscape perturbs the fitness in the local populations: increasing the fitness of some populations (e.g. when predators emigrate from a low prey-density cell), while decreasing the fitness of others (e.g. when prey disperse from a high predator-density cell). A change in the local fitness has a direct effect on the reproduction rates and hence on the size of the local population in the next generation. Therefore, this motion process brings the local populations to a state of high sensitivity to dispersal events. The migration of a single individual may cause new dispersal events in surrounding cells and hence may cause "avalanches" of dispersal. From this mechanism of dispersal emerge highly correlated regions of population density (figure 2f-g). The boundaries of these patches are, however, not well defined. Although this system is not purely deterministic, the complex patterns observed are evocative of spatiotemporal chaos in fluids (Cross and Hohenberg, 1993). The structure factors of the spatial population density in this regime indicate the absence of any characteristic length scale at which to describe those patterns. This is visible from the power-law shape of the structure factor when plotted on a log-log scale (figure 2i). The exponent of the observed power law has been computed for each value of $p_{motion}$ in this regime and the values are reported in table 1. All exponents in regime $II$ have value -2+$\varepsilon$, where $\varepsilon$ indicates a small deviation[1]. This result is quite remarkable as the exponent -2 is characteristic of self-similar systems. Indeed, all mean-field theories of systems of two coexisting phases in equilibrium (here low and high population density) yield an exponent -2 (Goldenfeld, 1992; Reichl, 1998). Nevertheless, we cannot guarantee that this behavior will be conserved in systems of size larger than the one investigated here, as changes in scaling regimes have been observed in other simulated and natural systems (Crawley and Harral, 2001; Allen and Holling, 2002; Pruessner and Jensen, 2002) .

Regime $III$ , of organized domains, corresponds to large fitness threshold values. Surprisingly, the spatial patterns emerging from this type of dispersal are highly structured (figure 2j-k). The boundaries separating regions of high and low densities are quite clear. In this regime, $p_{motion}$ is so high that almost every individual on the landscape has fitness inferior to the threshold. As a consequence, all individuals are in constant motion. Populations in each habitat are redistributed evenly amongst the cells of their neighborhood. Dispersal has thus a local homogenization effect. Therefore the reproduction process becomes locally predominant over the dispersal process in generating the patterns of population density. We believe that the structured regions of high and low densities may thus be spatial analogues of the common temporal predator-prey cycles.

---

[1] Some of the differences in the values of the exponents from -2 to -3 can be accounted for by a crossover scaling ansatz (Filotas, in preparation).

| | Disordered | | | Complex | | | | Organized |
|---|---|---|---|---|---|---|---|---|
| $p_{motion}$ | 0.3 | 0.4 | 0.5 | 0.6 | 0.7 | 0.8 | 0.9 | 1.0 |
| **Exponent** | -0.85 ±0.03 | -1.40 ±0.13 | -1.68 ±0.07 | -1.92 ±0.06 | -1.95 ±0.04 | -2.15 ±0.05 | -2.37 ±0.11 | -3.14 ±0.23 |

**Table 1.** Exponent of the power law for the structure factor as a function of $p_{motion}$ in the model with community-driven dispersal. Every exponent is obtained by finding the slope of a linear fit on a log-log graph of the structure factor (averaged over time and over 100 simulation runs), all give $r^2$>0.99.

The significant distinction between the spatial structures produced in the complex regime *II* and in the organized regime *III* can be easily measured by the structure factor. Once again the structure factor obeys a power law (figure 2m), however the exponent is now near -3 (table 1). While the scaling region leading to this exponent is narrow, it is large enough to measure the distinct exponent and to corroborate the correspondence between the qualitative change in patterns and the quantitative measurement of their structure factor.The exponent -3 is consistent with Porod's law (Porod, 1982), a theory in condensed matter physics stating that the structure factors of two-dimensional systems containing two well-separated phases (again the low and high population density) have a $1/k^3$ behavior.

Even if patterns of all sizes are present in this regime, too, their smooth shape indicates the absence of complex spatiotemporal patterns.

The power laws arising in the structure factors (regimes *II* and *III*) are seen to be unable to include points of very low wavenumber. This limitation is a consequence of the finite size of the lattice. The modeled landscape has size $L$=128, and thus spatial correlations cannot extend beyond this scale. We can test this suggestion by simulating the model on larger and smaller lattices. Figure 3 compares the structure factors obtained at $p_{\text{motion}}$ =0.7 for landscapes of size $L$=64, $L$=128 and $L$=256. We notice that the power law is improved on large length scales with the increase of the lattice size. Given this result, we have every reason to believe that for even larger systems the scaling region will be enhanced in the regime we have examined. Because of limited computer power we cannot, however, be sure that other phenomena could not appear in larger systems.

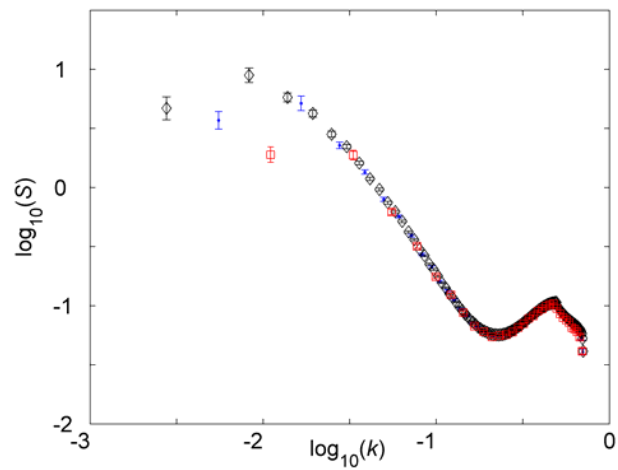

**Figure 3.** Log-log representation of the structure factor for the model with community-driven dispersal for three different lattice sizes: $L$=256 (diamond), $L$=128 (dot) and $L$=64 (square), for $p_{\text{motion}}$ = 0.7.

**Temporal analysis**

The temporal behavior of the model can be analyzed through the dynamics of $\rho_i(t)$ (figure 2d-h-l). As a first observation, we note that the average density of the prey, $\rho_v(t)$, decreases monotonically with $p_{motion}$, while the average predator density, $\rho_p(t)$, is almost constant. We suppose that this effect arises from the fact that, for low values of $p_{motion}$, predator and prey are mostly restricted to their original habitat giving rise to communities of even abundance throughout the landscape. The low predator abundance in each of these communities is an advantage for the prey which depends on the constant external resource for growth, and hence the prey's average population density stays high. As $p_{motion}$ increases, predator and prey become more mobile. The chase-escape motion leads to regions where the prey (predator) population is abundant and the predator (prey) population is small. In the regions where the prey is abundant, the predator population will tend to increase, while in the low prey abundance regions, the predator population will be diminished. On average, these two effects cancel out and explain the lack of change in the predator population density with $p_{motion}$. On the other hand, the prey are more sensitive to the predator's presence and the decrease of the predator population in certain regions is not enough to compensate for the negative effect of the increase of the predator population on the prey in the other regions.

Moreover, it is of interest to note that the temporal dynamics of the global variable $\rho_i(t)$ corroborates the spatial analysis of the dynamics. Indeed, the three-regime dynamics we have described is also evident from the variation of the average density $\rho_i(t)$ with respect to $p_{motion}$. In the disordered regime *I*, the temporal evolution of $\rho_i(t)$ oscillates only slightly around a mean value (figure 2d), while in the complex regime $II$, these oscillations develop into large fluctuations (figure 2h). These extreme variations can be explained by the increase of correlation which synchronizes the oscillatory dynamics of local populations over large areas.  In the organized regime $III$, domains are so well partitioned that they behave independently. This produces out-of-phase dynamics that cancel each other in the computation of the average density $\rho_i(t)$. Hence the fluctuations of $\rho_i(t)$ are reduced when regime $III$ is attained (figure 2i).  This statistical stabilization which reduces global predator-prey cycles is a common phenomenon which has been reported in other models such as the spatial Lotka-Volterra and the spatial Rosenzweig-MacArthur model (Jansen and de Roos, 2000) as well as in a spatial three-species competition model by Durrett and Levin (1998). To gauge the change in the amplitude of the fluctuations we show a graph of the standard deviation of the average density, the time-independent $\sigma(\rho_i)$ (equation 9), as a function of $p_{motion}$ (figure 4a).

$$\sigma(\rho_i) = \sqrt{\left\langle \rho_i^2 \right\rangle - \left\langle \rho_i \right\rangle^2} \qquad (9)$$

This figure shows the resemblance of this dynamics with that of a phase transition. It should be noted, however, that for the investigated system size, we were unable to observe the sharp peak characteristic of a phase transition.

We have also investigated the robustness of the dynamics against modification of the prey's functional response. We replaced the ratio-dependent Holling type $II$ response (equation 1) with a ratio-dependent Holling type $III$ (sigmoid) as well as a general ratio-dependent exponential response:

$$f_{III} = \eta_i \frac{(R/N(x,y,t))^2}{1+(R/N(x,y,t))^2} \qquad (10)$$

$$f_e = 1 - \exp(-\eta_i R/N(x,y,t)) \qquad (11)$$

We find that the general dynamics is unaltered under different functional responses, but the boundaries between the three regimes change. However, for each of these scenarios we recover similar behavior for the standard deviation of the average density (figure 4). The existence of three different spatiotemporal regimes of dynamics is therefore unaffected by the foraging properties chosen for the model.

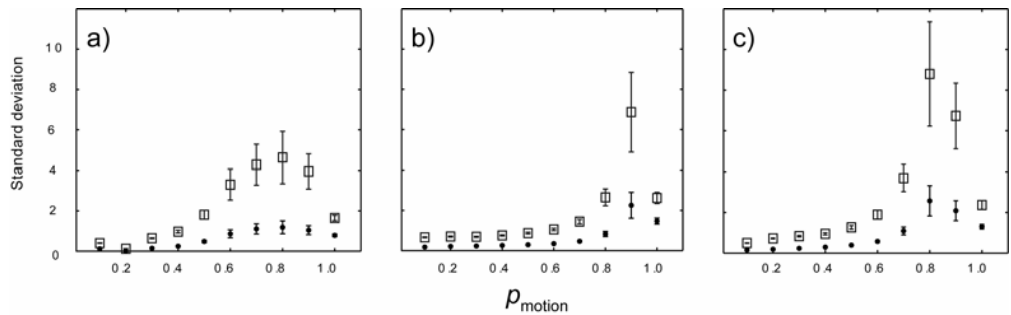

**Figure 4.** Standard deviation of the average population density $\rho$ for the model with community-driven dispersal as a function of $p_{motion}$ for the prey (square) and for the predator (dot). Three ratio dependent functional responses are represented: a.) Holling type $II$, b.) Holling type $III$ (sigmoidal) and c.) exponential.

## Impact of the scaling parameter

We have also investigated the variation of the three regimes with the scaling parameter $c$. Recall that $c$ varies inversely with the degree of the species sensitivity to the local biotic conditions. It hence defines the variability in species fitness in a community. When $c$ is set to a large value, for example above $c=0.4$, the fitness of every species is almost identical to 0.5. Hence, the fitness of a species is not affected by its local community and becomes independent of its position in the landscape. Therefore, the dispersal process cannot produce any increase or decrease of an individual's fitness. The species at every point on the lattice reproduce more or less at the same rate, which precludes any spatial patterns of population density to emerge. Thus the dynamics stays in regime *I* regardless of changes in $p_{motion}$ (figure 5).

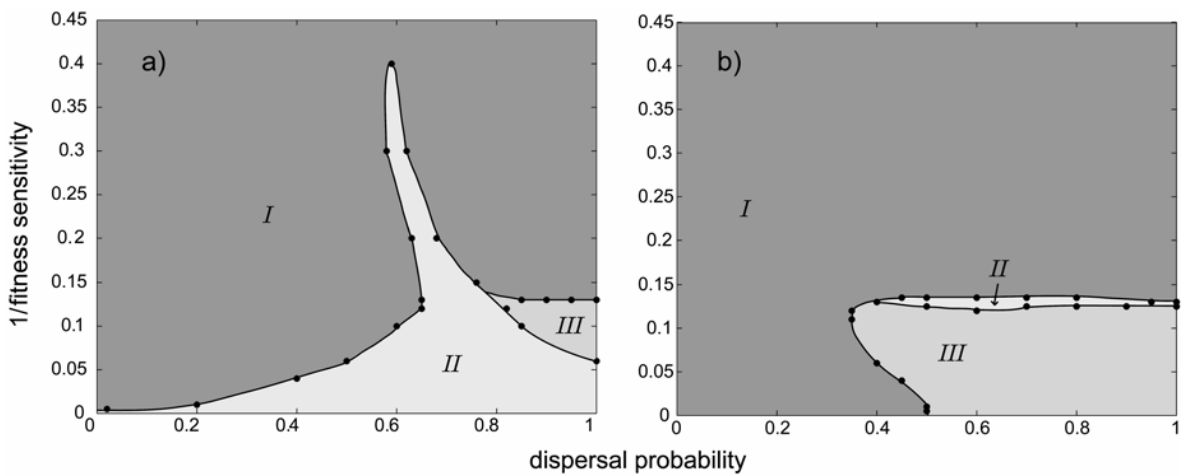

**Figure 5.** Dynamical regimes for a) community-driven dispersal and b) density-independent dispersal. The phase diagrams are expressed as functions of the level of fitness sensitivity (effect of scaling parameter *c*) and the dispersal probability ($p_{motion}$ in the community-driven case and $p_{ind}$ in the density-independent case). Points drawn on the diagram have been computed through simulations while the position of the phase boundaries has been deduced. Estimated 0.05 and 0.1 error on the dispersal probability should be considered in figure 5a and figure 5b respectively. As before, regime *I*, *II* and *III* correspond respectively to the disordered, complex and organized regime.

At the opposite end of the spectrum, a small *c* produces large variations in the species' fitness. This means that the fitness is very sensitive to changes in the local population sizes. Therefore, dispersal events, even of the smallest amplitude, modify considerably the local fitness and hence the local population density. As a result, at every $p_{motion}$ this dynamics generates complex spatial patterns and evolves in regime $II$ (see figure 5 at $c =$ 0.01).

Figure 5a depicts the model's behavior with changes in $p_{motion}$ and in $c$. The emergence of regime $II$, which corresponds to complex spatiotemporal patterns, is what distinguishes community-driven dispersal from density-independent dispersal. Therefore, in an ecosystem where predators and prey will usually have very distinct responses to their local environment (corresponding to $c$ in the range ~ (0.005, 0.1)), we expect complex population dynamics to be one of the possible outcomes of community-driven dispersal.

**4.2 Density-independent dispersal**

During dispersal controlled by a density-independent rate $p_{ind}$, a fixed proportion of individuals leave each cell of the landscape at each iteration of the model. The motion is independent of local fitness, and, as a result, individuals that are perfectly "happy" with their local biotic conditions may be forced to move out of their habitat in an artificial manner. Therefore, the main difference between this type of dispersal and the previously described community-driven motion, is that in each generation, each cell sees its population transformed by migration flow. Every population is obliged to participate in the dispersal process. As the dispersal probability $p_{ind}$ rises from 0 to 1, the participation of each population in the dispersal process increases in a linear fashion.

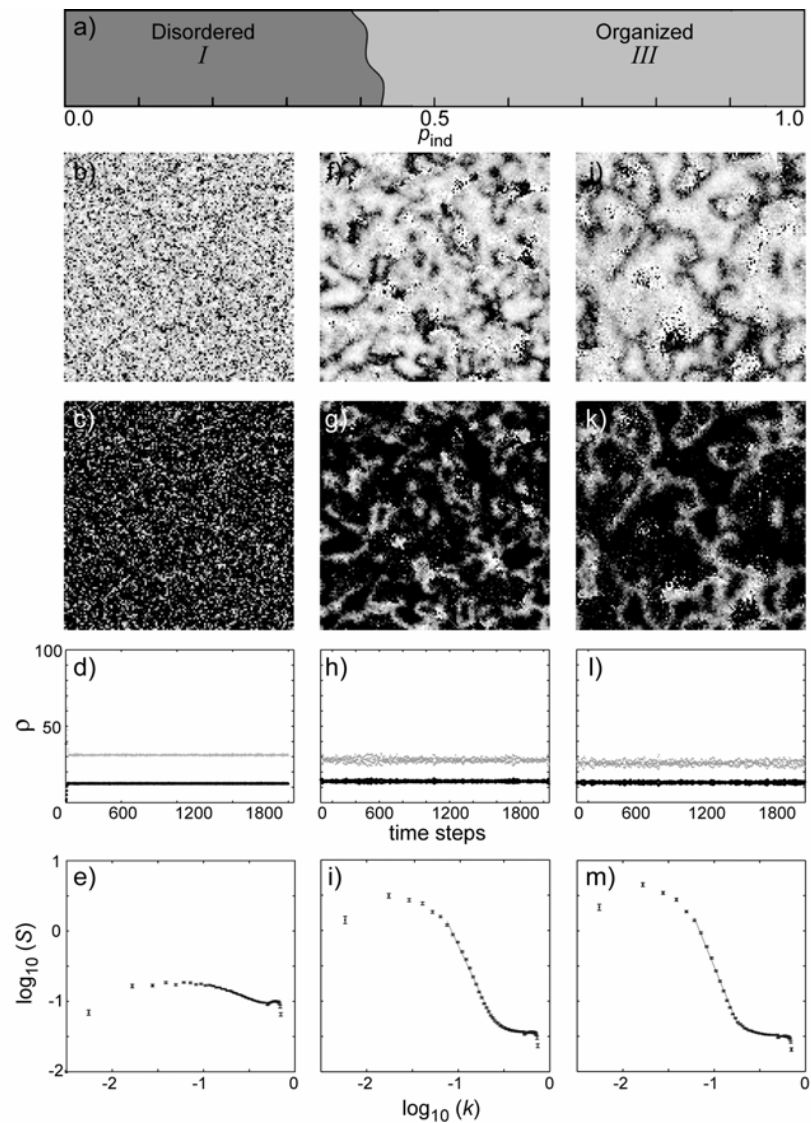

**Figure 6.** a.) Schematic representation of the two dominant dynamical phases for density-independent dispersal: *I* - disordered and *III* - organized domain (equivalent to figure 5b between *c* = 0.05 and *c* = 0.1). b.) Snapshot of the spatial prey density $D_v$ at $p_{ind}$ = 0.1. c.) Snapshot of the spatial predator density $D_p$ at $p_{ind}$ = 0.1. d.) Temporal evolution of the average density for the prey $\rho_v$ (gray) and for the predator $\rho_p$ (black) at $p_{ind}$ = 0.1. e.) Log-log plot of the structure factor of the spatial prey density at $p_{ind}$ = 0.1. (f-g-h-i) same as (b-c-d-e) but at $p_{ind}$ = 0.7. (j-k-l-m) same as (b-c-d-e) but at $p_{ind}$ = 1.0. All figures are obtained at *c* = 0.06.

The spatial patterns produced under this dynamics are consistent with this linear augmentation of the migration rate (figure 6a). For example, consider again the case $c$=0.06. When $p_{ind}$ is small, the interactions between the cells are weak, and no spatial structures are apparent (figure 6b-c). As $p_{ind}$ is increased, correlations in population density are induced by greater dispersal in the landscape, and small and definite patterns become visible (figure 6f-g). As $p_{ind}$ is increased further, those patterns develop into highly organized domains but conserve the same explicit profile (figure 6j-k). In fact, the spatial correlation increases in a smooth manner without any abrupt modifications in the structure factors (figure 6e,i,m). Note that the cases $p_{motion}$ =1.0 (figure 2j-m) and $p_{ind}$ =1.0 (figure 6j-m) are statistically identical because in both scenarios every individual continually disperses. Accordingly, the structure factor for $p_{ind}$=1.0 generates the same power law of exponent -3 as we find for $p_{motion}$=1.0. The exponents of the power laws for the cases $p_{ind}$<1.0 and c = 0.06 are reported in table 2. It is seen that the exponents increase continually from -2.5 at $p_{ind}$ = 0.5 (as soon as patterns are noticeable) to -3 at $p_{ind}$=1.0. This dynamics is therefore equivalent to going smoothly from regime *I* to regime *III* without passing through the spatiotemporal complex phase.

| | Disordered | | | | | Organized | | | | |
|---|---|---|---|---|---|---|---|---|---|---|
| $p_{ind}$ | 0.1 | 0.2 | 0.3 | 0.4 | 0.5 | 0.6 | 0.7 | 0.8 | 0.9 | 1.0 |
| **Exponent** | -0.46 ±0.04 | -1.16 ±0.07 | -1.74 ±0.04 | -2.13 ±0.10 | -2.41 ±0.18 | -2.66 ±0.11 | -2.80 ±0.04 | -2.99 ±0.04 | -3.07 ±0.09 | -3.15 ±0.08 |

**Table 2.** Exponent of the power law for the structure factor as a function of $p_{ind}$ in the model with density-independent dispersal. Every exponent is obtained by finding the slope of a linear fit on a log-log graph of the structure factor (averaged over time and over 100 simulation runs), all give $r^2$>0.99.

One should note that the value of the power-law exponent is not the only factor considered when assessing the nature of the dynamical regime as a function of dispersal. First, the entire shape of the structure factor should be taken into account. For example, the value of the exponents for $p_{ind}$=0.3 and $p_{ind}$=0.4 is very close to -2 (table 2), and one could be tempted to argue that they are part of a complex regime. On the other hand, their structure factors (not shown) have a weak $k$-dependence and therefore the power law does not span as many decades as the one obtained in the presence of complex spatiotemporal patterns. This indicates that the dynamics of the model at $p_{ind}$=0.3 and $p_{ind}$=0.4 seems to be somewhere between a disordered state and a highly-organized one, but should not be confused with the complex phase.

Second, the absence of the complex regime also appears in the variation of the average density $\rho_i(t)$ (figure 6 d-h-l and figure 7). Density-independent dispersal does not generate large fluctuations of the global population size as correlations cancel out between regions of domain organization.

Moreover, in the community-driven model, the average prey density decreases with $p_{motion}$ (figure 2 d-h-l) while this effect is not observed in the density-independent version (figure 6 d-h-l) where the prey density stays almost constant with the variation of $p_{ind}$. This is another consequence of the density-independent motion rule. Predators and prey are allowed to move regardless of their condition in the community, and this enhances the chance of encounters between the two species. As a result, prey highly suited to their community, that remained isolated from their predators in the community-dependent case, become more vulnerable in the density-independent counterpart, and their population density diminishes.

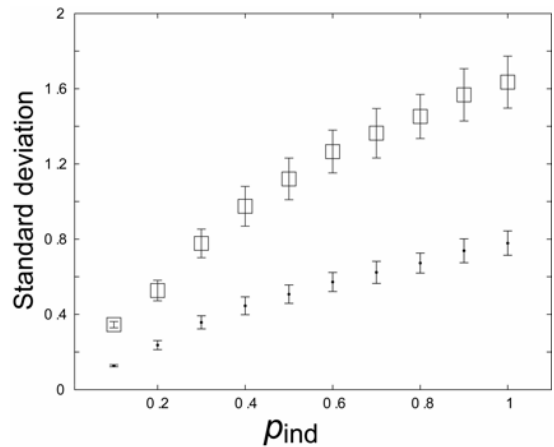

**Figure 7.** Standard deviation of the average population density $\rho$ for the model with density-independent dispersal as a function of $p_{ind}$ for the prey (square) and for the predator (dot). Note the different vertical scale from figure 4.

The impact of the scaling parameter $c$ on the density-independent dispersal model is shown in figure (5b). For large values of c, the dynamics remains in regime *I*. Spatial patterns do not emerge in a population of poor fitness variability. For low values of c (between 0 and 0.1) the dynamics is similar to that described earlier for c = 0.06: a smooth transition between a disordered state (regime *I*) to a highly organized state (regime *III*). In the region bounded by c = 0.11 and c=0.14, a narrow complex regime emerges. We suspect that for the given values of c, there must be a threshold in the population sizes, around which the fitness fluctuates rapidly to values above and below 0.5. The fitness of the local populations hence becomes quite sensitive to migration events, and as a result, regions of high and low population densities develop in a fractal fashion across the landscape. This complex regime is not found for values of $c$ below 0.11.

We should mention at this point that the boundaries separating the regimes in the phase diagram for the density-independent scenario (figure 5b) were harder to deduce than for the community-driven model. The reason is, that the variations of the spatiotemporal patterns, and hence of the structure factors, with dispersal probability occur gradually in the density-independent case with no sudden changes. The boundaries in the density-independent case could therefore represent crossovers between different types of behavior.

There is a notable difference between the phase diagrams of the community-driven and density-independent models. While both scenarios allow complex spatiotemporal patterns to emerge, these two complex regimes do not develop at the same level of fitness sensitivity. An important distinction is that the region of high fitness sensitivity (low $c$ values) in the phase diagram is occupied by the complex regime *II* in the community-driven case but is dominated by the organized regime *III* when dispersal is density-independent. Therefore, the density-independent model is unable to predict self-similar complex patterns of population density at a level of community sensitivity that we expect to find in natural ecosystems.

**6. Discussion and conclusion**

In this paper, we have presented a simple spatial predator-prey model. The model is based on a general hypothesis regarding species interactions, foraging and reproduction. The innovative feature of this model is that it includes the idea that dispersal is dependent on the local community. While it is known that dispersal is a way for individuals to escape communities for which they are poorly adapted, to our knowledge no spatial model so far has employed community-driven dispersal.

We found an interesting spatiotemporal dynamics markedly different from the one obtained using simple density-independent dispersal. This difference manifests itself through the appearance of a large complex regime, which occurs when species are particularly sensitive to the local environment. We expect biotic conditions in natural ecosystems to indeed have a great effect on species life history, and therefore complex population dynamics should be considered as one of the possible outcomes of community-driven dispersal.

The complex regime is caused by the motion rule, which depends on a tolerance threshold, and hence brings the local populations to a state of extreme sensitivity to dispersal events. A single migration event may thus propagate from one cell to another in a chain of dispersal. The resulting dynamic phase diagram is dominated by a complex (*II*) and a disordered (*I*) phase region, with a small organized phase region (*III*) for large dispersion threshold and intermediate fitness sensitivity, as shown in figure 5a. This threshold-based dynamical process cannot develop fully under density-independent dispersal, in which case every population participates equally in the dispersion process, irrespective of its condition in the local community. The result is a dynamic phase diagram dominated by the disordered (*I*) and organized (*III*) phases, with only a very narrow complex phase region (*II*) for intermediate fitness sensitivity, as shown in figure 5b.

Emergence of complex or chaotic spatiotemporal patterns is a much discussed topic of the past decades and has been observed in numerous spatial predator-prey models (Segel and Jackson, 1972; Hassell et al. 1991; Solé et al., 1992; Pascual, 1993; Wilson et al., 1993; Hassell et al. 1994; Bascompte et Solé, 1995; Gurney et al., 1998; Savil and Hogeweg, 1999; Sherratt, 2001; Biktashev et al. 2004; Marozov et al. 2004; Li et al. 2005; Marozov et al. 2006). Moreover, criticality in ecosystems undergoing phase like transition which results in the absence of characteristic spatial scale in patterns, has been reported in other studies (Solé et Manrubia, 1995; Malamud, 1998; Kizaki and Katori, 1999; Guichard, 2003; Pascual et Guichard, 2005). It is probable that self-organization is a common phenomenon in models based on growth-inhibition or recovery-disturbance processes and does not depend on the specific model details. On the other hand, not all self-organized patterns are alike, and different dispersal rules, as we have shown, can lead to different types of emerging patterns and hence have different ecological implications.

The conclusion of our study is therefore twofold. First, we demonstrate the relevance of studying spatial models, in which condition-dependent dispersal strategies are incorporated, since such dispersal strategies are common in nature (Bowler and Benton, 2005) and are likely to cause non-trivial dynamics. Second, we emphasize the need for comprehensive investigations on the relation between dispersal processes and spatial patterns. The structure factor provides significant information about the spatial structure and the scales of emerging patterns. We suggest that this method or similar spatial-correlation based techniques (Bjørnstad et al., 1999; Medvinsky et al., 2002; Morozov et al., 2006) are necessary for detailed comparison to be drawn between models varying in their dispersal strategy.

**Acknowledgments**

Funding for this research was provided by the Natural Sciences and Engineering Research Council of Canada (NSERC) and le Fond Québécois de la Recherche sur la Nature et les Technologies. We are thankful to the Réseau Québécois de Calcul de Haute Performance (RQCHP) for providing computational resources. Work at Florida State University was supported in part by U.S. National Science Foundation Grant Nos. DMR-0240078 and DMR-0444051. E. Filotas would like to thank V. Tabard-Cossa for his help on figure editing.

**References**

Abrams, P.A., Ginzburg, L.R., 2000. The nature of predation: prey dependent, ratio dependent or neither? Trends Ecol. Evol. 15, 337-341.

Allen, C.R., Holling, C. S., 2002. Cross-scale structure and scale breaks in ecosystems and other complex systems. Ecosystems 5, 315-318.

Amarasekare, P., 1998. Interactions between local dynamics and dispersal: insights form single species models. Theor. Pop. Biol. 53, 44-59.

Amarasekare, P., 2004. The role of density-dependent dispersal in source-sink dynamics. J. Theor. Biol. 226, 159-168.

Ariew A., Lewontin, R.C., 2004. The confusions of fitness. Brit. J. Phil. Sci. 55, 347-363.

Bascompte, J., Solé, R.V., 1994. Spatially induced bifurcations in single-species population dynamics. J. Anim. Ecol. 63, 256-264.

Bascompte, J., Solé, R.V., 1995. Rethinking complexity : modelling spatiotemporal dynamics in ecology. Trends Ecol. Evol. 9, 361-366.

Bascompte, J., Solé, R.V., 1998. Spatiotemporal patterns in nature. Trends Ecol. Evol. 13, 173.

Bascompte, J., Solé, R.V., Martinez, N., 1997. Population cycles and spatial patterns in snowshoe hares: an individual-oriented simulation. J. Theor. Biol. 187, 213-222.

Bilton, D.T., Freeland, J.R., Okamura, B., 2001. Dispersal in freshwater invertebrates. Annu. Rev. Ecol. Syst. 32, 159-181.

Biktashev, V.N., Brindley, J., Holden, A.V., Tsyganov, M.A., 2004. Pursuit-evasion predator-prey waves in two spatial dimensions. Chaos 14, 988-994.

Bjørnstad, O.N., Ims, R.A., Lambin, X., 1999. Spatial population dynamics: analyzing patterns and processes of population synchrony. Trends Ecol. Evol. 14, 427 - 432.

Blasius, B., Huppert, A., Stone, L., 1999. Complex dynamics and phase synchronization in spatially extended ecological systems. Nature 399, 354-359.

Bowler, D.E., Benton, T.G., 2005. Causes and consequences of animal dispersal strategies: relating individual behavior to spatial dynamics. Biol. Rev. 80, 205-225.

Christensen, K., di Collobiano, S.A., Hall, M., Jensen, H.J., 2002. Tangled nature: a model of evolutionary ecology. J. Theor. Biol. 216, 73-84.

di Collobiano, S.A., Christensen, K., Jensen, H.J., 2003. The tangled-nature model as an evolving quasi-species model. J. Phys. A: Math. Gen. 36, 883-891.

Crawley, M.J., Harral, J.E., 2001. Scale dependence in plant biodiversity. Science 291, 864-868.

Cross, M.C., Hohenberg, P.C, 1993. Pattern formation outside of equilibrium. Rev. Mod. Phys. 65, 851-1112.

Doebeli, M., Ruxton, D.G., 1998. Stabilization through spatial pattern formation in metapopulations with long-range dispersal. Proc. R. Soc. Lond. B 265, 1325-1332.

Durrett, R., Levin, S., 1998. Spatial aspects of interspecific competition. Theor. Pop. Biol.

53, 30-43.

Eldredge, N., Gould, S.J., 1972. Punctuated equilibria: an alternative to phyletic gradualism. In: T.J.M. Schopf (Editor), Models in paleobiology, Freeman, Cooper & Company, San Francisco, pp. 82-115.

Getz, W.M., 1984. Population dynamics: a resource per capita approach. J. Theor. Biol. 108, 623-644.

Goldenfeld, N., 1992. Lectures on phase transitions and the renormalization group. Addison-Wesley, Boston.

Guichard, F., Halpin, P.M., Allison, G.W., Lubchenco, J., Menge, B.A., 2003. Mussel disturbance dynamics: signatures of oceanographic forcing from local interactions. Am. Nat. 161, 889-904.

Gurney, W.S.C., Veitch, A.R., Cruickshank, I., McGeachin, G.,1998. Circles and spirals: population persistence in a spatially explicit predator-prey model. Ecology 79, 2516-2530.

Hall, M., Christensen, K., di Collobiano, S.A., Jensen, H.J., 2002. Time-dependent extinction rate and species abundance in a tangled-nature model of biological evolution. Phys. Rev. E 66, 011904.

Hanski, I., 1998. Metapopulation dynamics. Nature 396, 41-49.

Hassell, M.P., Comins, H.N., May, R.M., 1991. Spatial structure and chaos in insect population dynamics. Nature 353, 255-258.

Hassell, M.P., Comins, H.N., May, R.M., 1994. Species coexistence and self-organizing spatial dynamics. Nature 370, 290-292.

Hassell, M.P., Miramontes, O., Rohani, P., May, R., M., 1995. Appropriate formulations for dispersal in spatially structured models: comments on Bascompte & Solé. J. Anim. Ecol. 64, 662-664.

Hastings, A., 1993. Complex interactions between dispersal and dynamics: lessons from coupled logistic equations. Ecology 74, 1362-1372.

Hogeweg, P., 1988. Cellular Automata as a paradigm for ecological modeling. Applied Mathematics and Computation 27, 81-100.

Jackson, A.L., Ranta, E., Lundberg, P., Kaitala, V., Ruxton, D.G., 2004. Consumer-resource matching in a food chain when both predators and prey are free to move. Oikos 106, 445-450.

Jansen, V.A.A., de Roos, A.M., 2000. The role of space in reducing predator-prey cycles. In: U. Dieckmann, R. Law and J.A.J Metz (Editors), The Geometry of Ecological Interactions, Cambridge University Press, Cambridge, pp. 183-201.

Jensen, H.J., 2004. Emergence of species and punctuated equilibrium in the Tangled

Nature model of biological evolution. Physica A 340, 697-704.

Johst, K., Brandl, R., 1997. The effect of dispersal on local population dynamics. Ecol. Modell. 104, 87-101.

Johst, K., Schops, K., 2003. Persistence and conservation of a consumer-resource metapopulation with local overexploitation of resources. Biol. Cons. 109, 57-65.

Kean, J.M., Barlow, N.D., 2000. The effects of density-dependence and local dispersal in individual-based stochastic metapopulations. Oikos 88, 282-290.

Kendall, B.E., Bjørnstad, O.N., Bascompte, J., Keitt, T.H., Fagan, W.F., 2000. Dispersal, environmental correlation, and spatial synchrony in population dynamics. Am. Nat. 155, 628-636.

Kizaki, S., Katori, M., 1999. Analysis of canopy-gap structures of forests by Ising-Gibbs states – Equilibrium and scaling properties of real forests. J. Phys. Soc. Jpn. 68, 2553-2560.

Lawson, D., Jensen, H.J., 2006. The species-area relationship and evolution. J. Theor. Biol. 241, 590-600.

Li, Z.-z., Gao, M., Hui, C., Han, X.-z., Shi, H., 2005. Impact of predator prey pursuit and prey evasion on synchrony and spatial patterns in metapopulation. Ecol. Modell.185, 245-254.

Maguire, B. Jr., 1963. The passive dispersal of small aquatic organisms and their colonization of isolated bodies of water. Ecological Monographs 33, 161-185.

Malamud, B.D., Morein, G., Turcotte, D.L., 1998. Forest-fires : an example of self-organized critical behavior. Science 281, 1840-1842.

Marquet, P.A., 2000. Invariants, scaling laws, and ecological complexity. Science 289, 1487-1488.

Medvinsky, A.B., Petrovskii, S., Tikhonova, I.A., Molchow, H., Li, B.-L., 2002. Spatiotemporal complexity of plankton and fish dynamics. SIAM Review 44, 311-370.

Morozov, A., Petrovskii, S., Li, B.-L., 2004. Bifurcations and chaos in a predator-prey system with Allee effect. . Proc. R. Soc. Lond. B 271, 1407-1414.

Morozov, A., Petrovskii, S., Li, B.-L., 2006. Spatiotemporal complexity of patchy invasion in a predator-prey system with Allee effect. J. Theor. Biol. 238, 18-35.

Nathan, R., 2006. Long-distance dispersal of plants. Science 313, 786-788.

Packard, N.H., Wolfram, S., 1985. Two-dimensional cellular automata. J. Stat. Phys. 38, 901-946.

Pascual, M., 1993. Diffusion-induced chaos in spatial predator-prey system. Proc. R. Soc.

Lond. B 251, 1-7.

Pascual, M., Guichard, F., 2005. Criticality and disturbance in spatial ecological systems.Trends Ecol. Evol. 20, 88-95.

Platt, T., Denman, K.L., 1975. Spectral analysis in ecology. Annu. Rev. Ecol. Syst. 6, 189-210.

Porod, G., 1982. General Theory. In: O. Glatter and L. Kratky (Editors.), Small angle X-ray scattering, Academic Press, New York, pp. 17-51.

Press, W.H., Flannery, B.P., Teukolsky, S.A., Vetterling, W.T., 1992. Numerical Recipes in C: The art of scientific computing, Second Edition. Cambridge University Press, New York.

Pruessner, G., Jensen, H.J., 2002. Broken scaling in the forest fire model. Phys. Rev. E 65, 056707.

Ranta, E., Kaitala, V., 2000. Resource matching and population dynamics in a two-patch system. Oikos 91, 507-511.

Ranta, E., Kaitala, V., Lundberg, P., 1997. The spatial dimension in population fluctuations. Science 278, 1621-1623.

Reichl, L.E., 1998. A Modern Course in Statistical Physics. Second Edition. Wiley & Sons, New York.

Renshaw, E., Ford, E.D., 1984. The description of spatial patterns using two-dimensional spectral analysis. Vegetatio 56, 75-85.

Rikvold, P.A., 2006. Complex Behavior in Simple Models of Biological Coevolution. Submitted to Int. J. Mod. Phys. C. Preprint: arXiv:q-bio.PE/0609013.

Rikvold, P.A., 2007. Self-optimization, Community Stability, and Fluctuations in Two Individual-based Models of Biological Coevolution. J. Math. Biol. 55, 653-677.

Rikvold, P.A., Sevim, V., 2007. An Individual-based Predator-prey Model for Biological Coevolution: Fluctuations, Stability, and Community Structure. Phys. Rev. E 75, 051920.

Rikvold, P.A., Zia, R.K.P., 2003. Punctuated equilibria and 1/f noise in a biological coevolution model with individual-based dynamics. Phys. Rev. E 68, 031913.

Rohani, P., May, R.M., Hassell, M.P., 1996. Metapopulations and equilibrium stability: the effects of spatial structure. J. Theor. Biol. 181, 97-109.

Ruxton, D.G., 1996. Density-dependent migration and stability in a system of linked populations. Bull. Math. Biol. 58, 643-660.

Ruxton, D.G., Rohani, P., 1999. Fitness-dependent dispersal in metapopulations and its consequences for persistence and synchrony. J. Anim. Ecol. 68, 530-539.

Sæther, B.-E., Engen, S., Russell, L.,1999. Finite metapopulation models with density-dependent migration and stochastic local dynamics. Proc. R. Soc. Lond. B 266, 113-118.

Savil, N.J., Hogeweg, P., 1999. Competition and dispersal in predator-prey waves. Theor. Pop. Biol. 56, 243-263.

Segel, L.A., Jackson, J.L., 1972. Dissipative structure: an explanation and an ecological example. J. Theor. Biol. 37, 545-559.

Sevim, V., Rikvold, P.A., 2005. Effects of correlated interactions in a biological coevolution model with individual-based dynamics. J. Phys. A: Math. Gen. 38, 9475-9489.

Sherratt, J.A., 2001. Periodic travelling waves in cyclic predator-prey systems. Ecol. Lett. 4, 30-37.

Solé, R.V., Bascompte, J., Valls, J., 1992. Stability and complexity of spatially extended two-species competition. J. Theor. Biol. 159, 469-480.

Solé, R.V., Manrubia, S.C., 1995. Are rainforests self-organized in a critical state? J. Theor. Biol. 173, 31-40.

South, A., 1999. Dispersal in spatially explicit population models. Cons. Biol. 13, 1039-1046.

Wilson, W.G., de Roos, A.M., McCauley, E., 1993. Spatial instabilities within the diffusive

Lotka-Volterra system: individual-based simulation results. Theor. Pop. Biol. 43, 91-127.

Wootton, J.T., 2001. Local interactions predict large-scale pattern in empirically derived cellular automata. Nature 413, 841-844.

Ylikarjula, J., Alaja, S., Laakso, J., Tesar, D., 2000. Effects of patch number and dispersal patterns on population dynamics and synchrony. J. Theor. Biol. 207, 377-387.

Zia, R.K.P., Rikvold, P.A., 2004. Fluctuations and Correlations in an Individual-based Model of Biological Coevolution. J. Phys. A: Math. Gen. 37, 5135-5155.